\documentclass[prf,twocolumn,superscriptaddress]{revtex4-1}
\usepackage{amsmath,amssymb}
\usepackage{graphicx}
\usepackage{bm}
\usepackage{color}
\usepackage{hyperref}
\usepackage{float} 

\begin{document}

\title{Marangoni Fingering Instabilities in Oxidizing Liquid Metals}

\author{Keith D. Hillaire}
 \email{kdhillai@ncsu.edu}
\affiliation{Dept. of Physics, NC State University, Raleigh, NC}
\author{Michael D. Dickey}%
\affiliation{Dept. of Chemical and Biomolecular Engineering, NC State University, Raleigh, NC}
\author{Karen E. Daniels}%
\affiliation{Dept. of Physics, NC State University, Raleigh, NC}

\date{\today}

\begin{abstract}
Eutectic gallium-indium (EGaIn), a room-temperature liquid metal alloy, has the largest tension of any liquid at room temperature, and yet can nonetheless undergo fingering instabilities. This effect arises because, under an applied voltage, oxides deposit on the surface of the metal, which leads to a lowering of the interfacial tension, allowing spreading under gravity. Understanding the spreading dynamics of room temperature liquid metals is important for developing soft electronics and understanding fluid dynamics of liquids with extreme surface tensions. When the applied voltage or the oxidation rate becomes too high, the EGaIn undergoes fingering instabilities, including tip-splitting, which occur due to a Marangoni stress on the interface. Our experiments are performed with EGaIn droplets placed in an electrolyte (sodium hydroxide); by placing the EGaIn on copper electrodes, which EGaIn readily wets, we are able to control the initial width of EGaIn fingers, setting the initial conditions of the spreading. Two transitions are observed: (1) a minimum current density at which all fingers become unstable to narrower fingers; (2) a current density at which the wider fingers undergo a single splitting event into two narrower fingers. We present a phase diagram as a function of current density and initial finger width, and identify the minimum width below which the single tip-splitting does not occur. 
\end{abstract}

\maketitle

\section{\label{s:intro}Introduction}

Understanding fluid instabilities of liquid metals in the presence of electric fields is not only a fundamental fluid dynamics and electrochemistry problem, but also of importance for developing applications. While gallium-based room-temperature liquid metals show promise as electrical conductors for soft, flexible, tunable electronics \cite{dickey_stretchable_2017,li_electrochemically_2019,jin_stretchable_2015,markvicka_autonomously_2018,park_review_2013, zeng_fiber-based_2014} and soft robotics \cite{kim_soft_2013, rus_design_2015}, their  extraordinarily high interfacial tension \cite{dickey_eutectic_2008, gheribi_temperature_2019}  makes fabrication difficult. Furthermore, room temperature liquid metals could provide a convenient model system for improving our understanding of liquid metals in more extreme contexts, including 3D printing with metal powders \cite{sames_metallurgy_2016}, liquid metal batteries \cite{kelley_fluid_2018}, and molten aluminum casting \cite{grandfield_oxide_2014}.
One promising experimental system arises from the observation that the electrochemical oxidation of eutectic gallium-indium (EGaIn) dramatically lowers its surface tension 
\cite{eaker_liquid_2016, zhang_electrically_2019, wang_liquid_2019, wissman_field-controlled_2017}, even to the point of suppressing the Rayleigh-Plateau  instability in falling streams \cite{song_overcoming_2020}. Yet, when oxidation is used to lower the surface tension to promote spreading in droplets, the resulting flows exhibit fractal-like fingering instabilities
\cite{eaker_oxidation-mediated_2017}.

\begin{figure}
    \centering
    \includegraphics[width=\columnwidth]{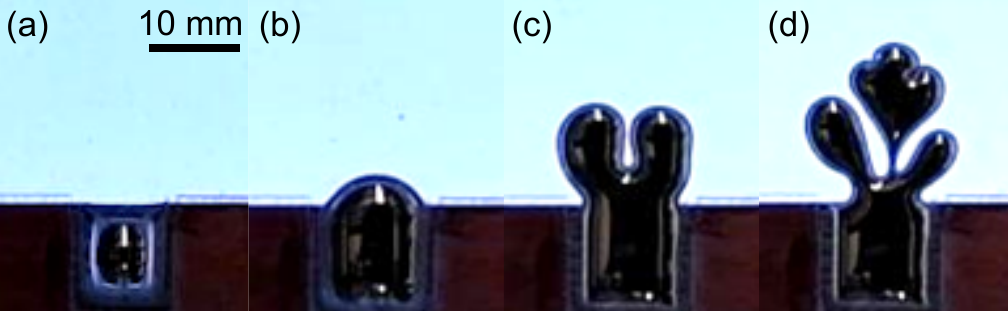}
    \caption{Examples of the EGaIn deposited on the copper electrode while not spreading (a) and the three spreading phases (b-d) that occur when the current density is high enough to induce spreading. If the current is too low, the interfacial tension too high, the hydrostatic pressure is not enough to press the EGaIn off of the copper electrode. At higher currents we see droplet-like spreading to a round finger (b), single tip-split spreading to two daughter fingers (c), and repeated splitting to fractal-like morphology (d).}
    \label{f:phases_examples}
\end{figure}

Gallium alloys have been the focus of much research over the last decade due to their low melting point and low toxicity \cite{surmann_voltammetric_2005}. In this paper, we perform experiments on eutectic gallium-indium (EGaIn), a room temperature liquid metal alloy comprised of 75.5\% gallium and 14.5\% indium by weight. EGaIn has the highest known surface tension of any room temperature liquid, and does not readily wet non-metallic materials, and so droplets of EGaIn typically maintain a highly spherical shape \cite{dickey_eutectic_2008} in the absence of surface oxidation. However, when a droplet within an electrolytic bath is exposed to an electric potential, the resulting oxidation decreases its surface tension to the point where the droplet undergoes a cascade of fingering instabilities as it spreads due to gravitational forces \cite{khan_giant_2014, eaker_oxidation-mediated_2017}. 

Possible mechanisms for the instability include viscous fingering \cite{saffman_penetration_1958}, 
electrokinetics \cite{melcher_electrohydrodynamics_1969,lin_instability_2004,gao_active_2019}
and Marangoni forces \cite{huppert_flow_1982}. Viscous fingering is ruled out because the intruding fluid (EGaIn) has a higher viscosity than the displaced electrolyte, although only by a factor of two.  Experiments by 
\citet{eaker_oxidation-mediated_2017} ruled out electrokinetics and electrohydrodynamics as the dominant mechanisms by placing electrodes directly above the droplet and an in-plane circle around the droplet, and found that the droplet spread similarly in both setups. The electrostatic forces were compared to the gravitational forces by measuring the angle at which their dynamics could be stalled by tilting the apparatus, and also found to be negligible. Because the electrochemistry is known to produce gallium oxide at the interface, Marangoni stress has been hypothesized as the mechanism for the instability. As with aqueous surfactants, each end of the oxide molecule has a preference for a particular side of a droplet's interface.

Marangoni fingering instabilities were first identified as tears of wine \cite{thomson_xlii_1855} in 1855. The instability is induced by a gradient in the surface tension, which may itself be caused by a gradient in temperature \cite{sur_steady-profile_2004}, solution concentration \cite{dukler_theory_2020}, or surfactant concentration \cite{warner_fingering_2004}. In each case, a gradient in interfacial tension causes a stress on the interface, leading a flow on the interface from lower interfacial tension to higher interfacial tension, so that any inhomogeneities are amplified into a finger-like pattern \cite{craster_dynamics_2009}. In thermally-driven Marangoni instabilities, \citet{sur_steady-profile_2004} observed a critical width (wavelength) of the fingers, below which the surface tension relaxed any gradients, suppressing finger-formation. Above this critical finger width, surface tension causes any perturbations along on the interface to grow, resulting in a fingering instability. 

In this paper, we aim to quantify the fingering instability in a model liquid metal system, and directly verify that Marangoni stresses at the interface give rise to the observed fingering instabilities.
We perform experiments on EGaIn placed in an electrolytic bath of 1 M NaOH, which has a pH of 14, which dissolves the excess oxide from the surface of the metal. Electrolytes also increase the conductivity of the bath, allowing an electrical current to be driven through the liquid metal and across the bath, controlling the oxidation rate of the liquid metal. The combination of control over oxidation and dissolution allows us to control oxide growth at the interface. The EGaIn eventually returns to its initial condition after dropping the voltage to below open circuit potential which electrochemically strips the oxygen from the interface, leaving only the bare metal.

Our experiments follow a method similar to that of \citet{sur_steady-profile_2004}, but setting an initial width $W_0$ for individual EGaIn fingers rather than creating a sinusoidal wavefront to determine which finger widths are the most stable. This design suppresses fluid exchange between fingers and thereby avoids the ``two balloon effect'' which otherwise drives the transfer of fluid from the smaller finger to the larger finger due to the difference in Laplace pressure \cite{merritt_pressure_1978}. We quantify the relationship between initial finger width $W_0$,  finger growth rate $\bar{v}$,  current density $J$, and surface tension $\gamma$ using the novel tensiometer design, as shown in Fig.~\ref{f:setupcartoon}, which utilizes top down images to determine $\gamma$. Although prior work has focused on experiments at fixed potential difference $\phi$ \cite{khan_giant_2014,eaker_oxidation-mediated_2017, song_overcoming_2020}, here we perform our experiments at constant current to fix a constant oxidation rate and thereby facilitate future modeling. We examine the behavior as a function of $J$ to make the findings area-independent.

We observe three distinct spreading behaviors, illustrated in Fig.~\ref{f:phases_examples}. At low $J$ (low oxidation rate), the fingers spread until they reach a steady state length, referred to as droplet-like spreading since this is equivalent to the pendant drop in a more conventional tensiometer geometry. At intermediate  $J$, we observe that wide fingers split into  narrower fingers, referred to as tip-splitting  by analogy with a similar instability observed in directional solidification \cite{utter_alternating_2001}. When $J$ is high enough (high oxide concentration), we observe that all fingers undergo repeated cycles of splitting, resulting in a fractal-like morphology over a limited range of length scales.

\section{\label{s:setup}Experimental Methods}

\begin{figure}
    \centering
    \includegraphics[width=\columnwidth]{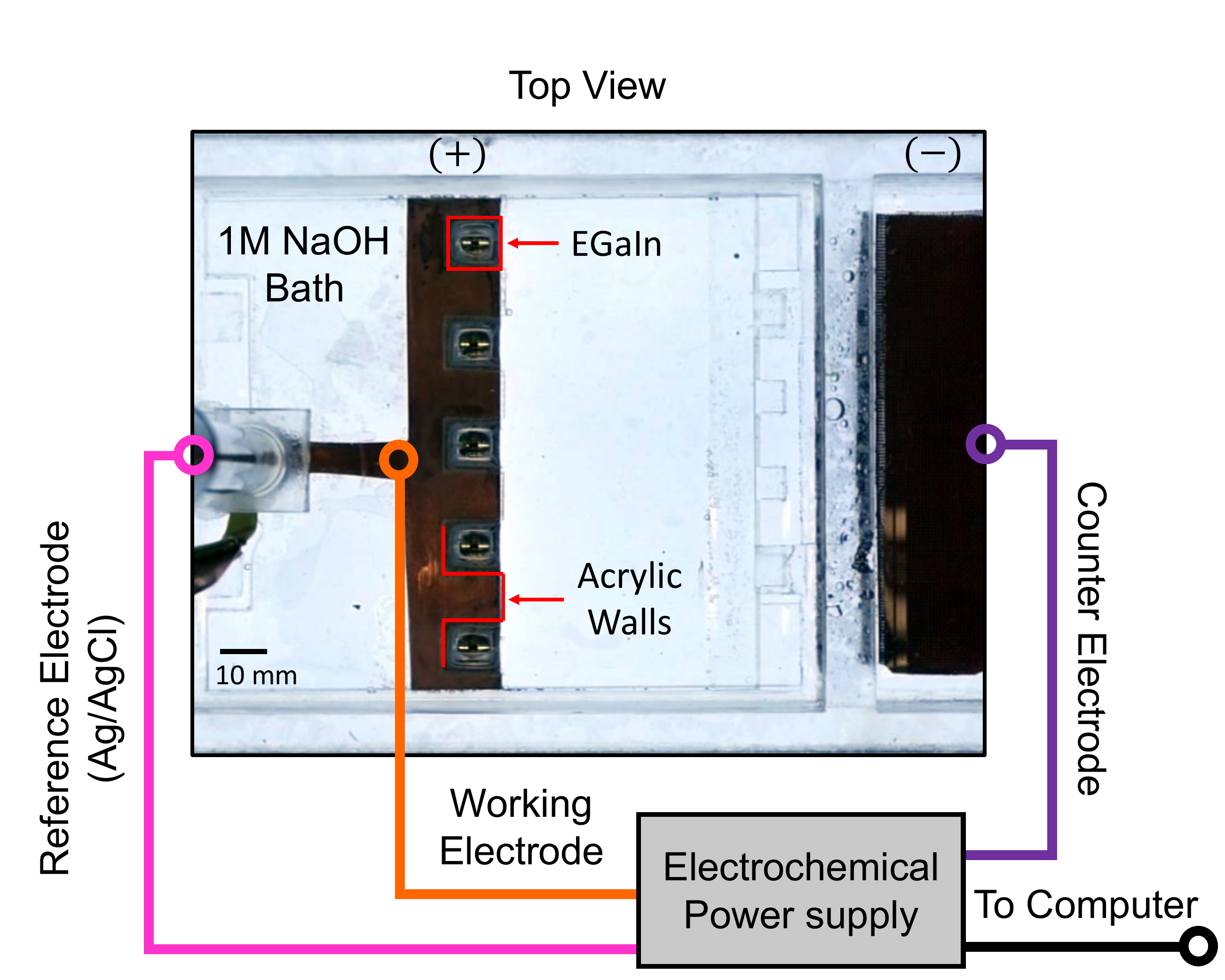}
    \caption{Schematic of the experimental setup. Working electrode is a copper sheet while the counter electrode is a copper mesh. Laser cut acrylic separates the initial EGaIn fingers.}
    \label{f:setupcartoon}
\end{figure}

\begin{figure*}
    \centering
    \includegraphics[width=0.8\linewidth]{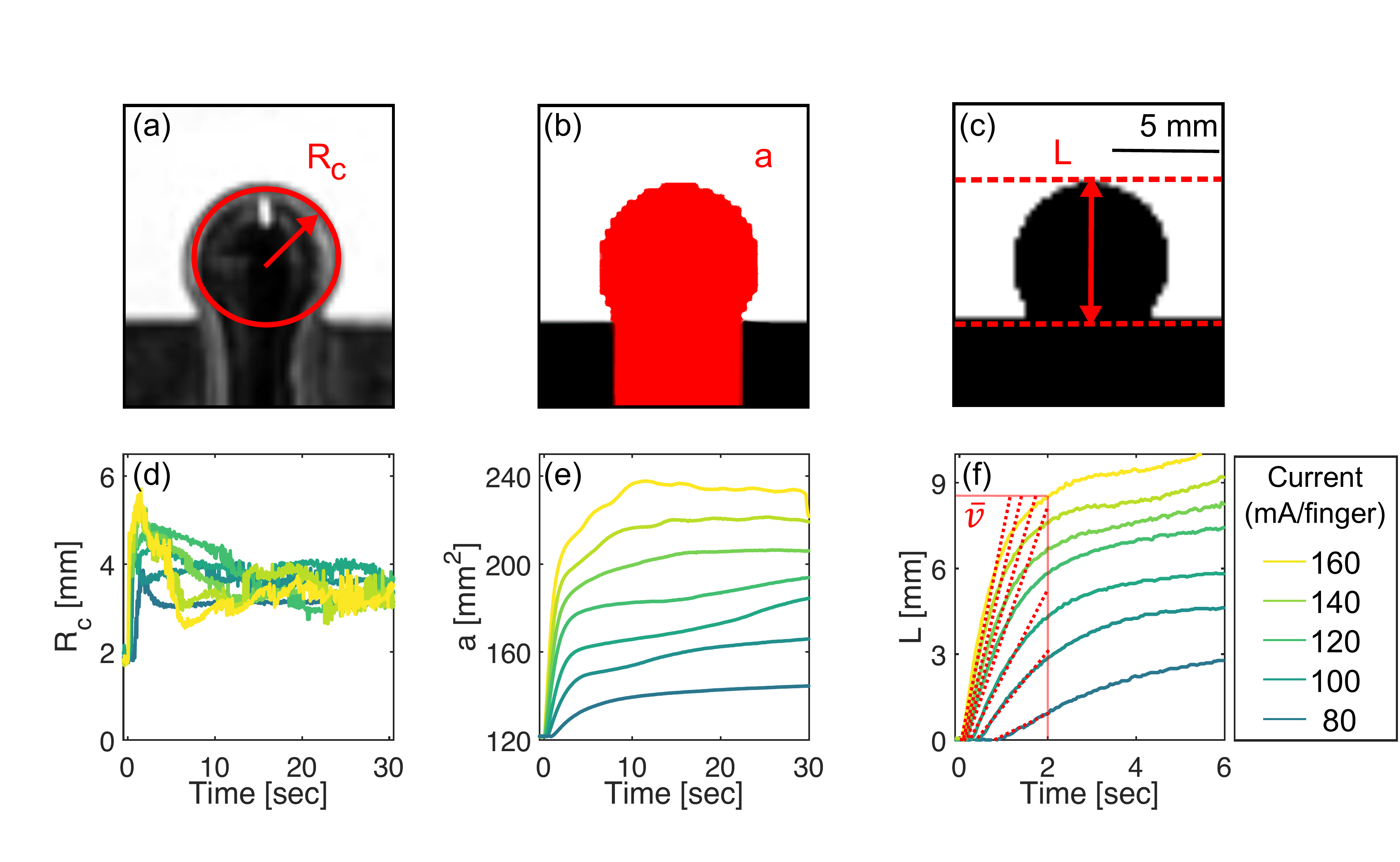}
    \caption{Sample image analysis techniques, shown for a single finger: (a) radius of curvature $R_c$, (b) cross-sectional area $a$, (c) finger length $L$. Plots (d-f) show sample dynamics over a subset of runs done using a $W_0 =7.2$ mm electrode, and a sample calculation of the the linearized growth rate $\bar{v}$.}
    \label{f:analysisexample}
\end{figure*}

\paragraph*{Apparatus:} 
We conducted experiments by applying an electric potential to five individual droplets of EGaIn immersed in an electrolytic bath, as shown in Fig.~\ref{f:setupcartoon}. 
We used a Keithley 2400 in a 3-electrode configuration, with a working, reference, and counter electrodes, so that it operates as a galvanostat. The working electrode is a 0.25 mm thick copper sheet and the counter electrode is 0.25 mm diameter copper wire mesh; the use of copper is based on its stability  in highly basic solutions and its wettability by EGaIn. The reference electrode is an Aldrich Ag/AgCl glass reference electrode, selected for its long term stability in a high pH environment.

The bath is constructed from laser cut, 6.35 mm thick, transparent, continuous cast acrylic, and holds 500 mL of 1M NaOH solution. As shown in Fig.~\ref{f:setupcartoon}, the bath holds all three electrodes at the same fixed locations for all experiments. The bottom acrylic plate of the bath contains three leveling screws, adjusted prior to each experimental run.
The working electrode is connected to the electrode plates from which five EGaIn droplets spread. The copper electrode plates are sandwiched between an acrylic base that snaps into the walls of the bath, and a patterned template of walls defines the initial width $W_0$. The walls are configured so that across all seven values of $W_0$ ($4.2$ to $10.2 \pm 0.1$ mm), each exposed copper patch has a surface area of  $120~\mathrm{mm}^2$. The choice of $W_0$ is set by the resolution of the laser cutter for small $W_0$, and by the width of the bath for large $W_0$, but covers the full regime over which the instabilities occur.

\paragraph*{Data collection:} 
The primary experimental dataset comprises 21 runs (each at a different fixed current) for each of the seven electrodes setting the initial $W_0$. We place an EGaIn droplet of volume $V = 200$~$\mu$L on each of the five copper pads. For each run, the selected current is applied for a duration of 60 seconds, 
sufficient time for the stable fingers to reach a steady-state and stop growing. The fixed current values range from $I =  0$ to 1 A (200 mA/finger). All runs for a given initial condition were performed on a single day, with a minimum of five minutes between experimental runs to ensure dissolution of the oxide. Data is averaged over the middle 3 fingers to exclude edge effects that occur on the outer fingers. In addition, we conduct several other types of runs to determine the dynamics on the surface of the EGaIn as it spreads, such as placing tracer particles on the surface of the EGaIn and growing a thick oxide shell so that we can use the reflection of light off the interface to measure the dissolution of the oxide.

\paragraph*{Image processing:}
We monitored the dynamics of the droplet shape and growth with a monochromatic Pixelink PL-A741 camera, with a resolution of 600$\times$600 pixels and at frames rates up to 23 Hz. All images used for image-processing were taken from above, with the sample backlit from below using an LED panel to provide a uniform light field.
From individual images, we characterized three properties (see Fig.\ref{f:analysisexample}a-c). The radius of curvature $R_c$ was determined using a Hough transform on the original image. To measure the droplet dimensions (cross-sectional area $a$, length $L$) the image is first binarized. We measured $a$ from the number of black pixels beyond the edge of the electrode, and $L$ from the furthest outward distance reached. The calibration from pixels to mm is done using the known width of the working electrode.
These three measurements underlie the calculation of several derived quantities. In \S\ref{s:tensiometer}, we will calculate surface tension from  $R_c$ and $a$ and the known droplet volume. We measure each finger's initial growth rate by fitting a line to $L(t)$, and calculate the current density from  $J = I/a$ for each finger.

\paragraph*{Oxidation and dissolution rate:}
Throughout the paper, we assume that the oxidation rate increases monotonically with $J$, as is typical in electrochemistry as $\log(J)$ is most often proportional to the chemical reaction rate and potential across the interface, as seen in the Sand equation, the Tafel equation, the Butler-Volmer equation, and the Caberra-Mott equation \cite{bard_electrochemical_2001, cabrera_theory_1949}. We estimate the dissolution rate of the oxide using two different methods, and find that it is approximately independent of $J$ and $\phi$ for a fixed molarity.

\section{\label{s:Results} Results}

\subsection{\label{s:surfaceflow} Direct observation of surface flows}

\begin{figure}
    \centering
    \includegraphics[width=\columnwidth]{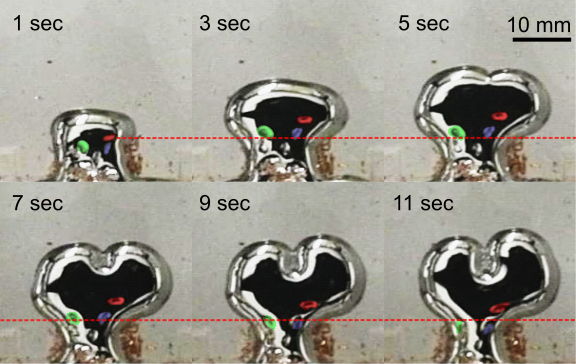}
    \caption{Time series of the growth of a finger, with copper tracer particles placed on the surface, undergoing a tip-splitting instability. The red arrows, all at the same starting positions, point to its updated location in that frame. Top row: tracers move with the bulk finger growth. Bottom row: several tracers move in the opposite direction of the finger growth.}
    \label{f:surfaceflow}
\end{figure}

To begin, we aim to examine the hypothesis that Marangoni stresses are present: these should drive the surface to move in the direction of the gradient in surface tension. As shown in Fig.~\ref{f:surfaceflow} and Video 1 (Supplemental Material), we place small copper particles onto the top surface of the EGaIn droplet and apply a fixed  oxidizing voltage. We observe that these tracer particles move outward (with the direction of finger growth) until the finger begins to split. 
During the splitting 
process, the tracer particles reverse direction and flow back toward the working electrode, against the growth direction of the finger itself.
If the fingering instability were due to viscous fingering, the tracer particles would have moved in the direction of finger growth due to inertia. While electrokinetic instabilities can cause fingering patterns in microchannels \cite{lin_instability_2004}, electro-osmotic flow requires an electric double layer in contact with a wall \cite{ramos_electrokinetics_2011}. In our system, the walls are not in contact with the moving liquid metal.
This leaves us to infer that tracer particle movement is due to the presence of a Marangoni stress, likely due to gradients in oxide concentration.

\subsection{\label{s:tensiometer} Measuring surface tension}

To measure surface tension $\gamma$, we adapt the non-invasive techniques used for standard, three-dimensional droplets \cite{liu_characterization_2012,khan_giant_2014} to apply them in a quasi-two-dimensional-top-down context. The experiment itself can be used as an in-situ tensiometer by considering the steady-state case when  Laplace pressure $P_{\mathcal L}$ is balanced by $P_{H}$. Solving for  $\gamma$, we find 
\begin{equation}
\gamma(h,R_c) = \Delta\rho gh  \left( \frac{1}{h} + \frac{1}{R_c} \right)^{-1}
\label{YLE}
\end{equation}
To calculate $h$ for use in this equation, we approximate the finger as a portion of a spherical cap  attached to a rectangular prism:
\begin{equation}
h \approx \frac{-a+(V+\sqrt{a^3+V^2})^{2/3}}{\pi^{1/2}(V+\sqrt{a^3+V^2})^{1/3}}
\label{SphericalCap}
\end{equation}
where  $a$ is the cross sectional area of the finger and $V$ is the known, deposited volume of the droplet.

\begin{figure}
    \centering
    \includegraphics[width=\columnwidth]{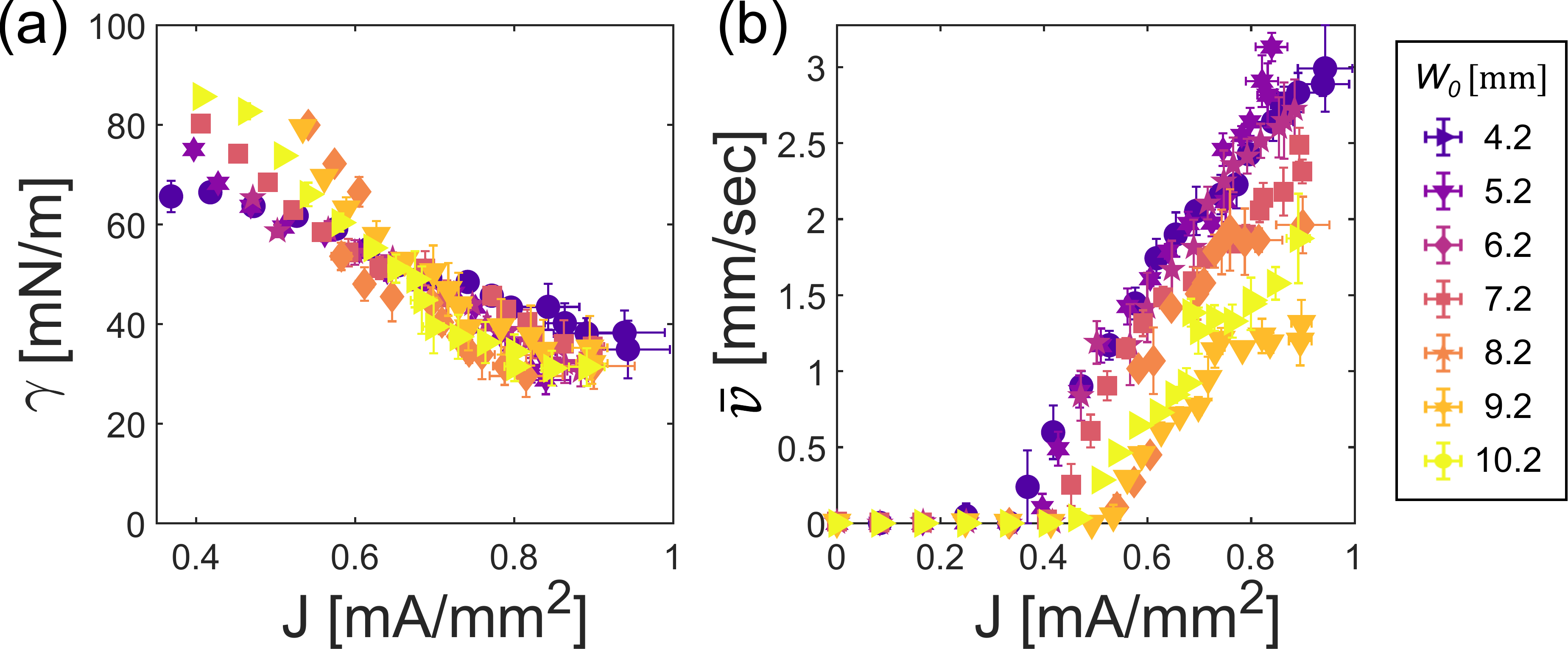}
    \caption{(a) $\gamma$ dependence on $J$. Error bars calculated by propagating the error of the directly measured values, $a$, $R_c$, and $V$, added in quadrature.
    (b) $\bar{v}$ dependence on $J$.}
    \label{f:ITandGrowth}
\end{figure}

This in-situ tensiometer allows us to calculate $\gamma$ throughout the regime in which round finger tips are observed. For each run, we calculate the average value of $\gamma$ for $10$~sec $< t < 30$~sec (as shown for example, in Fig. ~\ref{f:ITandGrowth}), corresponding to a period during which the fingers have reached a steady state and before the more unstable fingers pinch off the electrode, and also over the middle three (of five) fingers. In 
Fig.~\ref{f:ITandGrowth}a, we plot the calculated $\gamma$  as a function of $J$, and observe the expected result: that as the rate of creation of oxide increases, the surface tension decreases. We compared the $\gamma$ calculated using our tensiometer method to standard sessile drop method under the same conditions, as both a function of $J$ and $\phi$ and found both methods to produce the same results within error \cite{Song_OxidationMechanism}. We observe a subtle transition in $\gamma(J)$ which is reminiscent of similar measurements for insoluble surfactants on a liquid-liquid interface \citet{kaganer_structure_1999}.

\subsection{Finger growth and instability}

\begin{figure*}
    \centering
    \includegraphics[width=0.9\linewidth]{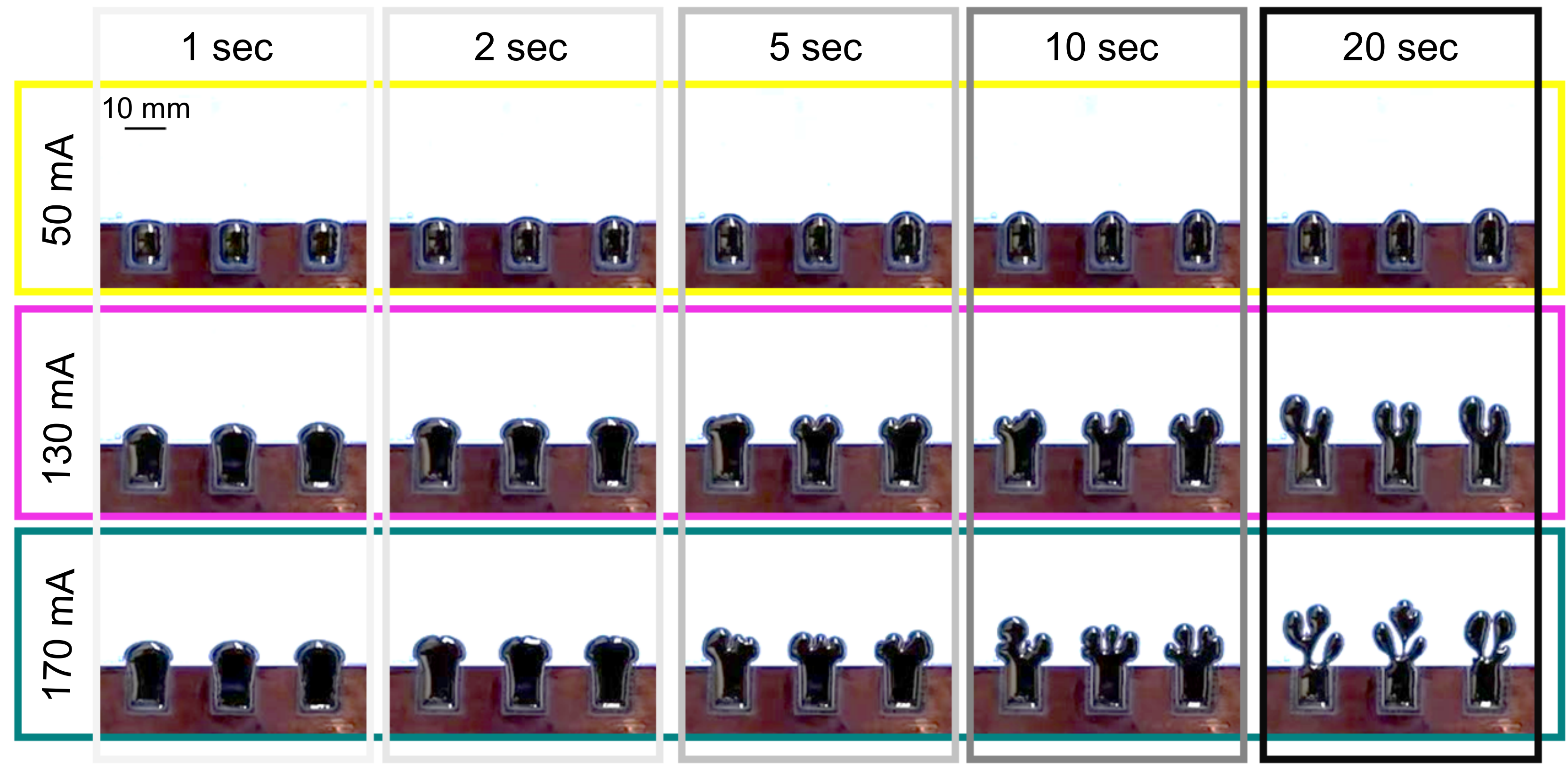}
    \caption{Differences in finger spreading behaviors over the first 20 seconds of spreading, for different values of current $I$. Frames from Supplemental Material videos V1, V2, and V3.}
    \label{f:TimeSeries}
\end{figure*}

Within each experimental run at fixed current and initial width $W_0$, any individual finger gradually grows in length as oxide deposit on its surface, $\gamma$ decreases, and gravitational forces push the fluid downward (outward). This process is plotted in Fig.~\ref{f:analysisexample}f, with both an initial linear-growth period and then a saturation to steady-state length. From such plots, we measure the initial tip speed $\bar{v}$ by fitting a line to the first 2 seconds of $L(t)$.

In Fig.~\ref{f:ITandGrowth}b, we observe that $\bar{v}$ increases with $J$, and decreases with $W_0$. The former observation coincides with an interpretation whereby a larger reaction rate (higher $J$) provides a faster increase in oxide concentration, and therefore a faster drop in surface tension and faster spreading.

Fig.~\ref{f:TimeSeries} shows that for low current $I$, the fingers do not grow as long as they do at higher currents, but maintain a round morphology. At intermediate $I$, the fingers undergo a single tip-split and those two fingers continue to grow. For sufficiently large $I$, the fingers continue to undergo tip-splitting instabilities until they form a fractal-like morphology.

\subsection{Phase Diagram}

\begin{figure}
    \centering
    \includegraphics[width=\columnwidth]{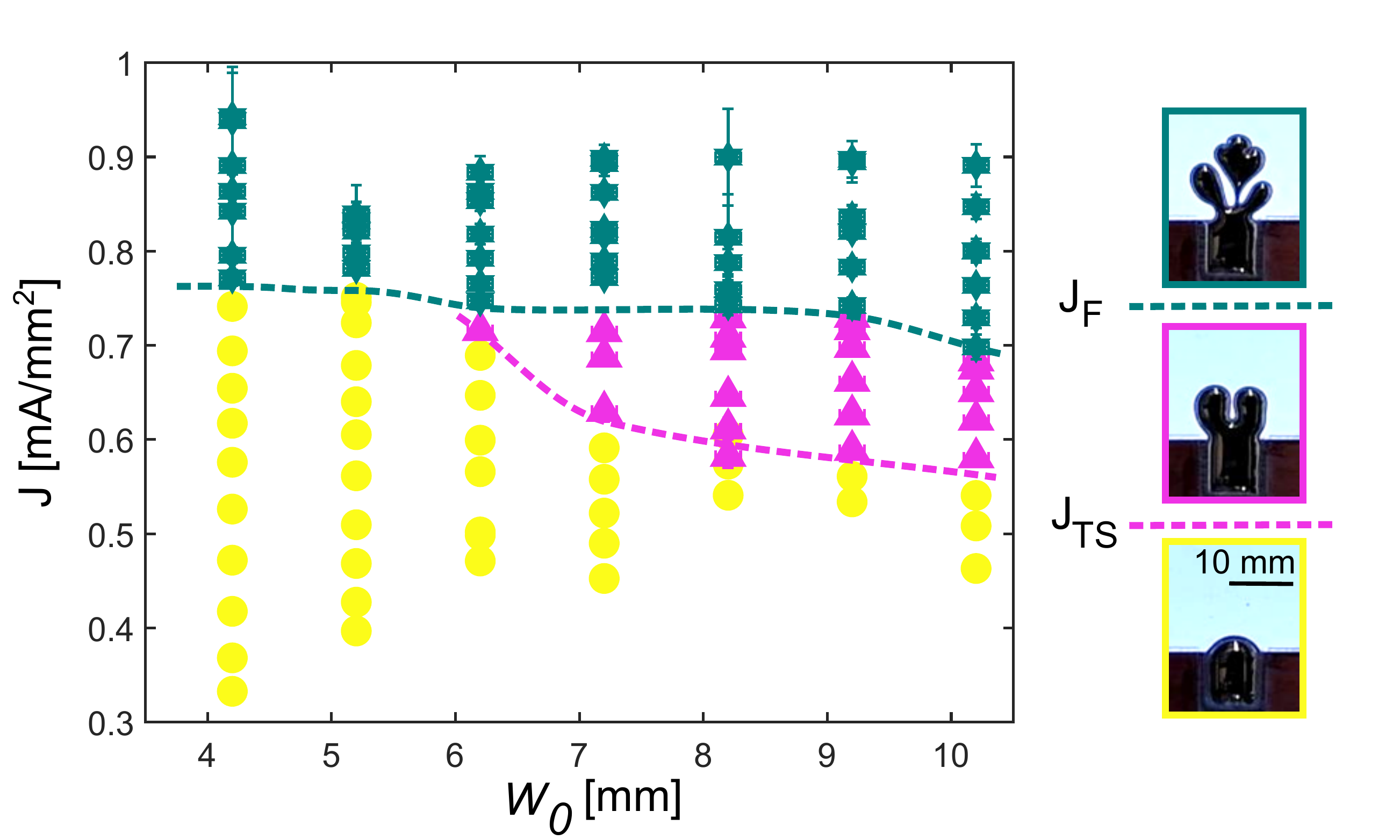}
    \caption{Phase diagram of the spreading behaviors as dependent on $J$ and $W_0$. Transition from round like morphology, yellow circles, to single tip-split, pink triangles, occurs with $J_{TS}$. Fingers with a current density higher than $J_F$ will undergo repeated splitting to a fractal-like morphology, green stars.}
    \label{f:PhasesJ}
\end{figure}

We can organize these observed spreading behaviors into a phase diagram, changing variables from $I$ to $J  \equiv I/a$. Fig.~\ref{f:PhasesJ} presents the observed behavior as a function of fixed $W_0$ and measured $J$. The behavior is classified according to whether they are droplet-like (round), tip-split, or fractal-like. Droplet-like fingers do not exhibit tip-splitting for the full 60 sec run, and tip-split fingers undergo only a single splitting.  Fingers may later undergo a pinch-off event, but that behavior is not considered in this paper.
For sufficiently low values of $J$, the EGaIn droplet remains confined to the working electrode, and these data points are not included in the phase diagram because we cannot measure $\gamma$ or $a$ within this region. 

Increasing $I$, and therefore $J$, allows the EGaIn to extend past the edge of the underlying copper plate, shown as the yellow circles in Fig.~\ref{f:PhasesJ}. The EGaIn forms a round finger with a shape similar to a pendant drop under gravity. Within this regime, the dominant forces on the fingers are due to hydrostatic pressure and Laplace pressure. As $J$ increases, the droplet decreases in $\gamma $ and increases $a$, until the droplet reaches a steady state. As shown in the first row of Fig.~\ref{f:TimeSeries}, the droplet-like fingers do not change shape over the time scale of the experiment.

 Fingers with $W_0 \ge$ 6.2 mm undergo a single tip-split event from one large parent finger to two smaller daughter fingers, as long as $J \le J_{TS}$, as shown in Fig.~\ref{f:PhasesJ} as the pink triangle. The 4.2 mm and 5.2 mm fingers did not undergo a single tip-split, rather their spreading behavior was droplet-like until $J \ge J_F$, the current at which the fingers undergo repeated splitting to a fractal-like morphology. Single tip-split fingers can reach a steady state, but typically will undergo an instability via the two balloon effect \cite{merritt_pressure_1978}. As one of the two fingers becomes too large, it eventually pinches off the rest of the EGaIn.

As $J$ increases, the fingers undergo a transition from droplet-like spreading to repeated splitting, which gives the fingers a fractal-like morphology, shown as the green stars in Fig.~\ref{f:PhasesJ}, similar to what was observed in \citet{eaker_oxidation-mediated_2017}. For the fingering instability to occur, the edge of the fingers must spread slower than perturbations that grow along the edge \cite{sur_steady-profile_2004}.
The fingers that undergo repeated tip splitting do not reach a steady state. When $J \ge J_F$, the fingers spread non-homogeneously until they have either completely spread across the system, or a section of the fingers pinch off from the rest of the finger.

Note that we have chosen $I$ as an independent parameter, rather the more typical $\phi$, and this leads to phase diagrams in which one axis is $J$. This choice is made because $J$ is proportional to the oxidation rate of the liquid metal, and since the fingers spread until they reach a steady state, the average oxide concentration increases monotonically with $J$. As shown in \ref{f:JoverV}, for $J < J_{TS}$, we found that $J$ increases monotonically with $\phi$, but this relationship breaks down once tip-splitting begins, the interface composition becomes non-homogeneous, and instabilities lead to more complicated surfaces for which a variety of other effects are present.

For completeness, the phase diagram is plotted as a function of $\phi$ and $I$ instead of $J$ are given in the Supplemental Material as  Figs.~\ref{f:PhasesPhi} and \ref{f:PhasesI}.

\section{\label{s:Discussion}Discussion}

We have observed that the fingers undergo three different spreading behaviors when the fingers are held at a constant current: round droplet-like, single tip-split, and repeated tip-splitting to a fractal-like morphology. We found the transition to the fractal-like morphology occurred when $J$ was greater than a critical $J$, $J \approx 0.7$~mA/mm$^2$, and fingers with initial widths of 6.2 mm or greater were observed to undergo a single tip-split event, occuring when the slower growing wider fingers to go unstable to the faster growing narrower fingers. We did not find $\phi$ to be constant at the transitions. Future experiments in which $\phi$ is held constant may change the characteristics of the spreading and/or the transitions between spreading behaviors.

The instabilities that lead to fingering are due to a competition between $\Delta P$ and Marangoni stress $\tau \propto \nabla\gamma$ \cite{bertozzi_undercompressive_1999}.
In other observations of tip-splitting due to Marangoni effect \cite{sur_steady-profile_2004}, the splitting occurs when $\tau$ becomes on the same order as other characteristic pressures. Here,  the hydrostatic pressure is given by
\begin{equation}
    P_{H} = \Delta\rho gh
\label{eq:hydrostatic}
\end{equation}
where $g$ is the acceleration due to gravity, $h$ is the average thickness of the finger, and $\Delta\rho$ is mass density difference between EGaIn and the 1 M NaOH.
In our system, by calculating $P_H$ at the transition to the fractal-like morphology, using the approximation for $h$ as given in Eq.~\ref{SphericalCap}, and $\Delta\rho$ as the difference in densities of EGaIn and 1 M NaOH, we observe that for $P_H \gtrsim 50$~Pa, perturbations form on the surface of the fingers, and ultimately go unstable. If we calculate $\tau$ by assuming a linear distribution of $\gamma$, where $\tau \approx (\gamma_0 - \gamma_{\phi})/L$, we find $\tau \approx 46$~Pa. If the maxima of the ripples, the new fingers, spread away from the bulk of the fingers faster than the bulk of the finger spreads, then this represents an instability \cite{sur_steady-profile_2004}. The daughter fingers would be narrower in width than the wider parent finger since the smaller fingers have a higher $\bar{v}$, as shown in Fig.~\ref{f:ITandGrowth}.

We used tracer particles, deposited on the surface of the EGaIn as shown in Fig.~\ref{f:surfaceflow}, to visually confirm the presence of the Marangoni stress. When the fingers grew without instabilities, the tracer particles moved in the same direction as the finger growth, and as instabilities formed, the tracer particles changed direction, flowing in the opposite direction of the finger growth.

The Marangoni instability, while it dominates, is not the only effect controlling these dynamics. Although we observed many tip splitting events, either single tip-splitting or repeated splitting to a fractal-like morphology, we did not observe any fingers merging while the fingers maintained contact with the working electrode. It is likely that the gallium oxide that forms on the surface of the fingers acts as a barrier, keeping two fingers from merging, or that the build up of excess charges on the interface may cause significant electrostatic repulsion.

Another challenge that remains is to identify which oxide specie(s) dominate the dynamics in each region of the phase diagram, Fig.~\ref{f:PhasesJ}. The reactions occurring on and around the EGaIn/bath interface are voltage dependent, and further experiments are needed to elucidate their interplay.

\section{Acknowledgements}

The authors would like to thank Minyung Song, William Llanos, Abolfazi Kiani, Sahar Nadimi, Jeffrey Wong, and Thomas Witelski for helpful discussions about the fluid dynamics and electrochemistry of EGaIn. We also like to thank the National Science Foundation for support under NSF DMR-1608097 (K.H. and K.E.D.) and NSF Grants CBET-1510772 and DMR-160897 (M.D.D.).

\bibliographystyle{apsrev}
\bibliography{eGaIn}

\newpage
\clearpage

\appendix
\renewcommand\thefigure{S\arabic{figure}}
\setcounter{figure}{0}

\section{Supplemental Material}

\subsection{Video captions}

\begin{description}
\item[V1]\url{https://www.youtube.com/watch?v=vMRojho7HlA} \\
EGaIn fingers undergoing droplet-like spreading: $I = 50$~mA/finger, $W_0 = 10.1$~mm, 200~$\mu$L/finger.

\item[V2] \url{https://www.youtube.com/watch?v=D3aTcCdqZi0} \\
EGaIn fingers  undergoing a single tip-splitting event: $I=130$~mA/finger, $W_0 = 10.1$~mm, 200~$\mu$L/finger.

\item[V3] \url{https://www.youtube.com/watch?v=14l4iVE1ScE} \\
EGaIn fingers  undergoing repeated tip-splitting events to a fractal-like morphology. $I=170$~mA/finger, $W_0 = 10.1$~mm, 200~$\mu$L/finger.

\item[V4] \url{https://www.youtube.com/watch?v=hHR0Gbcy1z8} \\ 
Tracer particles moving along the surface of EGaIn fingers as they undergo tip-splitting.

\end{description}

\subsection{Additional figures}

\includegraphics[width=0.9\columnwidth]{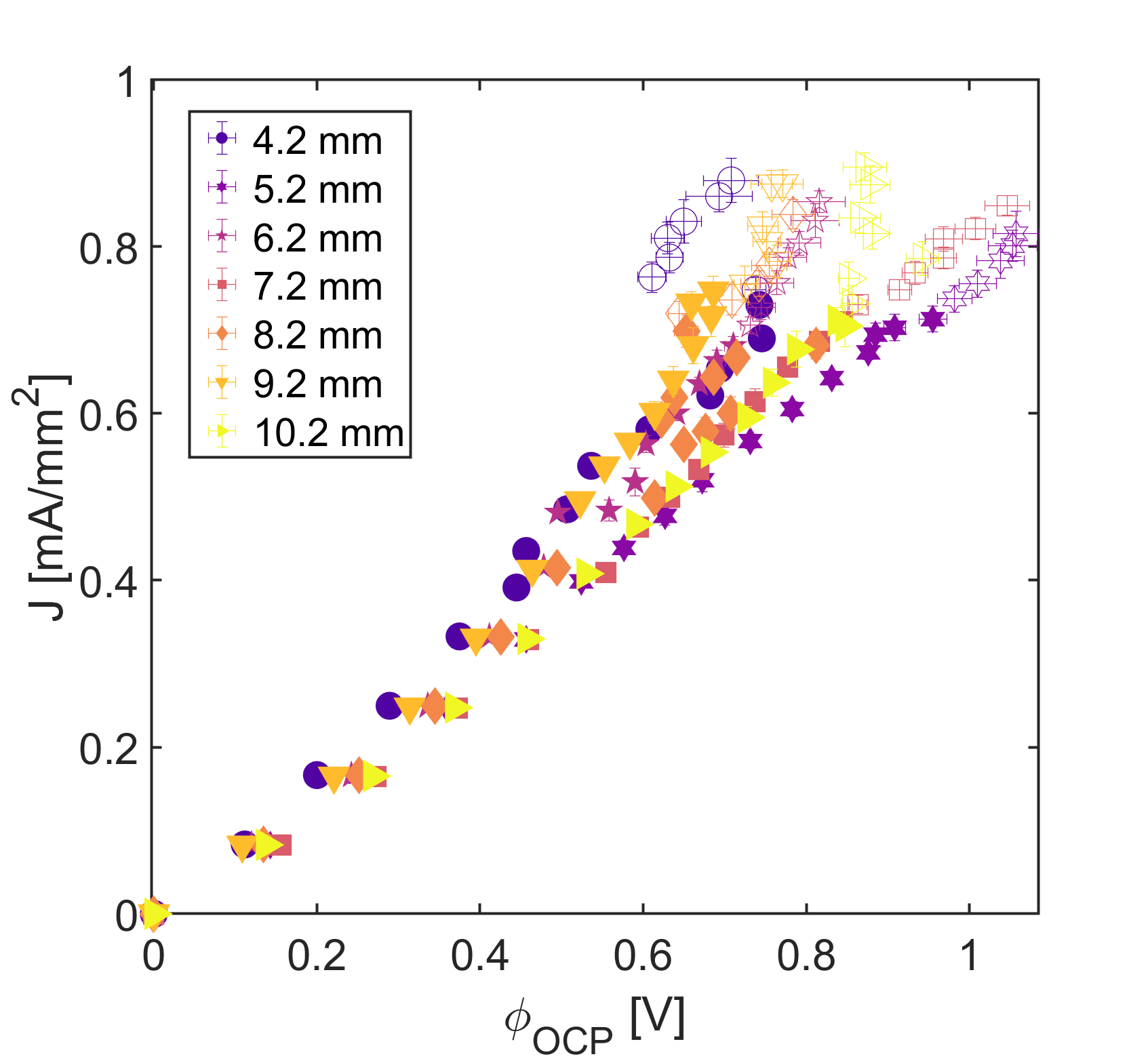}
\begin{figure}[h]
    \centering
    \caption{The relationship between current density and measured potential across the system. $J$ increases monotonically with $\phi$ until the instability forms, and there is a decrease in potential. Therefore the transitions in behavior should not depend on whether we hold the current or potential constant, though the growth rate and area of the fingers may change, the transition $J$ and $\phi$ should not.}
    \label{f:JoverV}
\end{figure}

\includegraphics[width=\columnwidth]{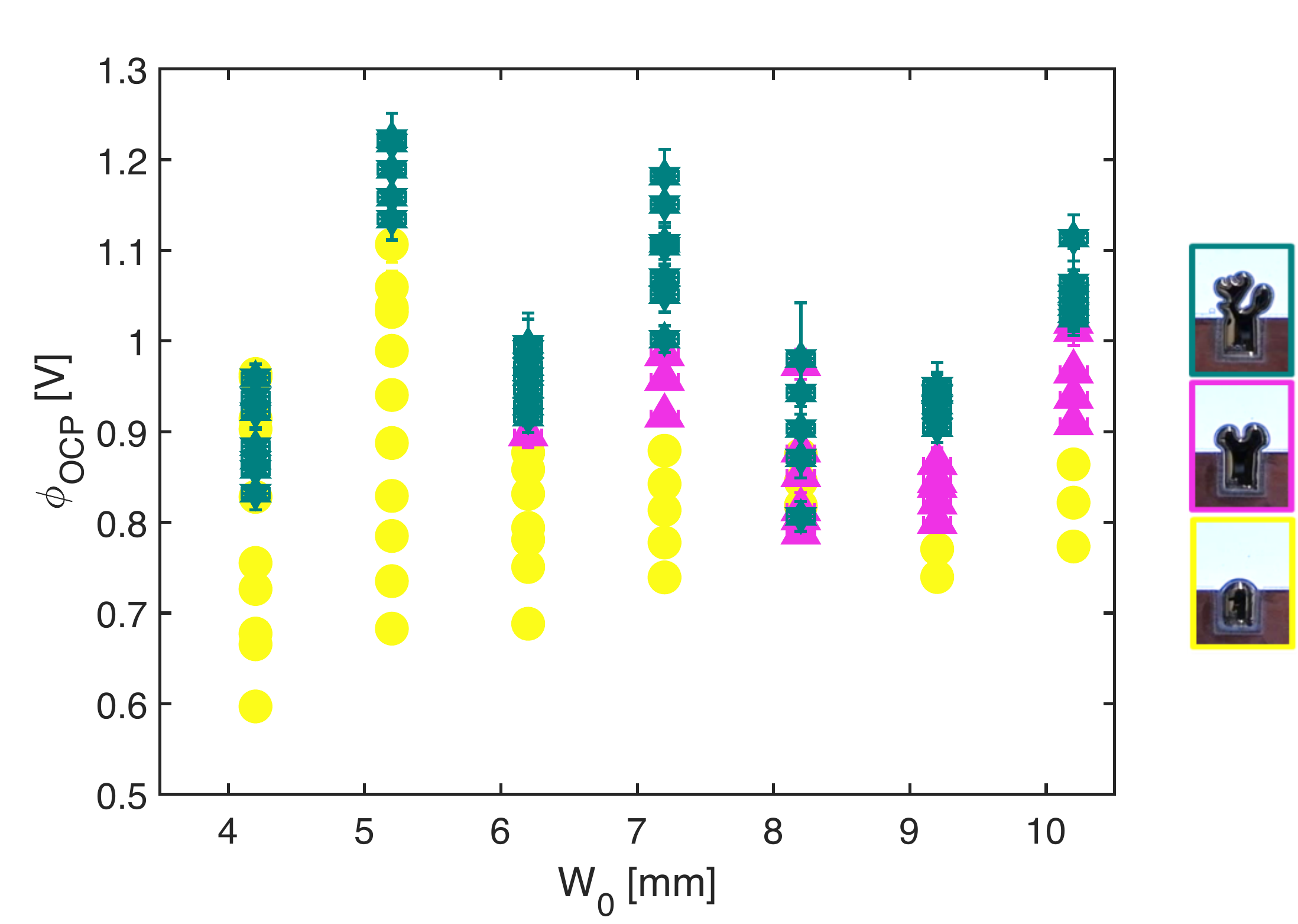}
\begin{figure}[h]
    \centering
        \caption{Phase diagram of the spreading behaviors, as a function of  $\phi$ and $W_0$, to compare with Fig.~\ref{f:PhasesJ}.}
    \label{f:PhasesPhi}
\end{figure}

\includegraphics[width=\columnwidth]{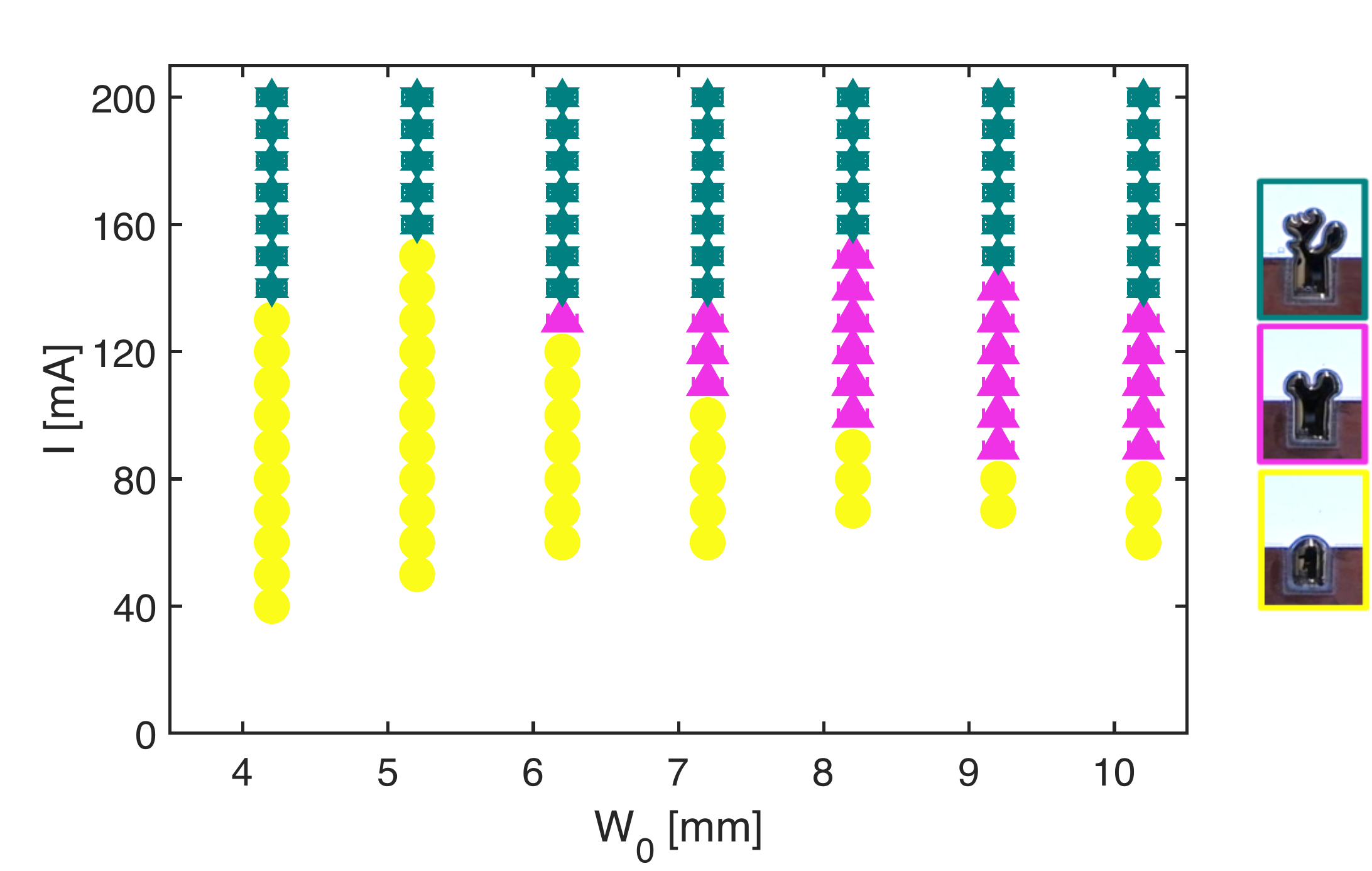}
\begin{figure}[h]
    \centering
        \caption{Phase diagram of the spreading behaviors, as a function of $I$ and $W_0$, to  compare with Fig.~\ref{f:PhasesJ}. Although the data points are evenly spaced in $I$, the transition into the fractal-like regime is less uniform. }
    \label{f:PhasesI}
\end{figure}

\end{document}